\documentclass[conference, 10pt, twocolumn]{IEEEtran}
\usepackage{cite}
\usepackage{amsmath,amssymb,amsfonts}
\usepackage{algorithmic}
\usepackage{graphicx}
\usepackage{textcomp}
\usepackage{xcolor}
\usepackage[nolist]{acronym}
\usepackage{comment}
\usepackage{mathrsfs,bm} 
\usepackage{scalerel}
\usepackage[caption=false]{subfig}
\usepackage{tikz}
\usepackage{pgfplots}
\usepackage{tabu}
\usepackage{mathtools}
\mathtoolsset{showonlyrefs}

\usepackage[
  hidelinks,
  bookmarks=false,
  pdfborder={0 0 0}
]{hyperref}

\hypersetup{colorlinks=true,linkcolor=black, citecolor=black, linktocpage}

\DeclareMathOperator{\diag}{diag}

\newcommand{\ee}{\mathrm{e}}
\newcommand{\mypi}{\mbox{\Large $\bm{\pi}$}}

\definecolor{pant24PeachFuzz}{RGB}{255,190,152}

\newcommand{\fc}{\textcolor{black}}
\usepackage{xcolor}
\usepackage{soul}
\usepackage{svg}

\IEEEoverridecommandlockouts
\begin{document}

\title{OTFS-IDMA: An Unsourced Multiple Access Scheme for Doubly-Dispersive Channels}

\author{Davide~Bergamasco, Federico~Clazzer, and Paolo~Casari
\thanks{D.~Bergamasco and F.~Clazzer are with the Inst.\ of Communications and Navigation, German Aerospace Center (DLR), Wessling, Germany (e-mail: \{davide.bergamasco, federico.clazzer\}@dlr.de).}%
\thanks{ D.~Bergamasco and P.~Casari are with DISI, University of Trento,  Italy (e-mail: \{davide.bergamasco, paolo.casari\}@unitn.it).}%
}

\maketitle
\thispagestyle{empty}

\begin{abstract}
We present an \ac{UMAC} scheme tailored to high-mobility wireless channels. The proposed construction is based on \ac{OTFS} modulation and sparse \ac{IDMA} in the \ac{DD} domain. The receiver runs a compressive-sensing joint activity-detection and channel estimation process followed by a single-user decoder which harnesses multipath diversity via the \ac{MRC} principle. Numerical results show the potential of \ac{DD}-based uncoordinated schemes in the presence of double selectivity, while remarking the design tradeoffs and remaining challenges introduced by the proposed design.
\end{abstract}
\section{Introduction}
\acresetall

\ac{IoT} applications are often characterized by a large number of devices aiming to report short and sporadic messages to a common receiver~\cite{nguyen20216g}.
Distributed transmitters may be difficult to coordinate without incurring large system overhead or unbearable delays, especially in settings with a massive amount of transmitting nodes. Considering these issues, an appealing alternative is given by grant-free \ac{MAC} schemes.

Grant-free \ac{MAC} strategies have been extensively studied in the past in the context of \ac{RA}~\cite{LIVA24}. The recent introduction of the \ac{UMAC} framework~\cite{UMAC} provided an achievability bound on the energy efficiency for uncoordinated \ac{MAC} schemes. It also proved that recent \ac{RA} solutions are not optimal. Hence, the research community has tackled the problem and proposed several new protocols, e.g.~\cite{T-FOLD-ALOHA,SPARCS,Tensor,SIDMA,ODMA}, that are closing the gap with the random-coding performance provided in~\cite{UMAC}. The initial analysis of the \ac{UMAC} was carried out assuming a \Ac{GMAC}. Successive research extended the bound and developed practical protocols for the quasi-static Rayleigh fading channel~\cite{Fasura,ODMA}, where the signal of each user is received with a different channel gain. 

In this paper, we considered the channel to be doubly-dispersive, as is typical in high-mobility settings. Relevant examples are vehicular networks or multi-user services relying on \ac{LEO} constellations. In satellite-enabled \ac{IoT} systems, a similar channel representation, characterized by residual time-frequency offsets, holds true when users are not able to perfectly compensate for the Doppler distortion and for the transmission delay due to imprecise information about the satellite trajectory or their own location.

Doubly-dispersive channels introduce strong distortions on the transmitted signals, forcing the receiver to counteract these impairments via equalization in order to retrieve the transmitted messages. It is worth noting that the commonly adopted single-tap equalization of \ac{OFDM} signals is not effective.
    Large subcarrier spacing helps reduce the performance degradation due to \ac{ICI} at the price of lower spectral efficiency, as the relative overhead from the \ac{CP} becomes more significant.
Additionally, to enable reliable equalization, the receiver needs to keep an up-to-date channel estimate. Its time validity is limited due to the short coherence time, thus requiring high pilot overhead. In~\cite{UMA-OFDM}, a scheme based on massive MIMO is proposed to deal with these challenges. In this work, we take a different approach by considering the \ac{DD} representation of the channel, which shows longer time stability and, in many practical scenarios, can be effectively characterized by a few parameters (see~\cite{OTFSHanzo,OTFSbits1,OTFSbook,2011matz_book} for a more comprehensive discussion). A convenient way of exploiting the \ac{DD} channel representation is to embed the information directly in the same domain using, e.g., \ac{OTFS} modulation~\cite{OTFS_original}. 

\ac{OTFS} has been investigated in the general context of random access in~\cite{sinha2020otfs} and~\cite{mirri2024} while \ac{DD}-based activity detection has been studied in~\cite{mattu2024delay,mirri2024amp}. By way of contrast, the main contribution of this paper is to develop a full \ac{UMAC} scheme that employs \ac{OTFS} to cope with doubly-dispersive channels. Different from previous \ac{DD}-based contributions, that focus on the design of the activity detection mechanisms, we develop and merge a second protocol step that enables the detection and decoding of the uncoordinated users messages. To achieve competitive performance, we expand some key ideas developed in the context of grant-free schemes for the \ac{GMAC} and adapt them to \ac{OTFS} modulation as well as to the additional challenges introduced by the doubly-dispersive model. \fc{Our implementation takes advantage from the principles of sparse \ac{IDMA} and \ac{ODMA} derived in~\cite{SIDMA} and~\cite{ODMA} respectively, where we adapted both the transmitter and receiver operations to work in the \ac{DD} domain over time-frequency selective channels. We remark the similarities between the \ac{IDMA} and \ac{ODMA} strategies. 
\Ac{IDMA} has been originally introduced in~\cite{IDMA}, while the its application to the \ac{UMAC} setting has been adopted in~\cite{SIDMA} and more recently in~\cite{SBIDMA}.}


The rest of the paper is organized as follows: In Section~\ref{sec:sys_mod}, we describe the \ac{UMAC} problem, the considered channel model, and the transmitter/receiver architectures. In Section~\ref{sec:numres}, we present numerical results on the performance of the proposed scheme, and we discuss design choices toghether with the role of the main parameters. Finally, in Section~\ref{sec:conc} we draw conclusions and propose future directions.

\section{System model}
\label{sec:sys_mod}
We consider a multi-user random access system, where we call $K$ the total number of users and $K_a$ the subset of active users. Each of the $K_a$ nodes transmits a $b$~bits message, mapped into $n$ complex channel uses through a common codebook $\mathcal{C}$. Every codeword $\bm{x}_k \in \mathcal{C} \subset \mathbb{C}^n$ fulfills the power constraint $||\bm{x}_k||_2^2\leq nP$. We consider a doubly-dispersive channel model, where the received signal $\bm{y}(t)$ takes the form:
\begin{equation}
\label{eq:channel}
\bm{y}(t) = \sum\limits_{k=1}^{K_a}\sum\limits_{p=1}^{P_k}h_{k,p}\,\bm{x}_k(t-\tau_{k,p}) \, {\rm e}^{\,j2\pi \nu_{k,p}(t-\tau_{k,p})} + \bm{w}(t) \, ,
\end{equation}
where $P_k$ is the number of propagation paths for the signal of user $k$, while $h_{k,p}$, $\tau_{k,p}$ and $\nu_{k,p}$ represent the channel gain, the delay and the Doppler shift of the $p$-th path, respectively. We assume every channel path to be subject to Rayleigh fading, hence $h_{k,p}$ are i.i.d. samples from the complex Gaussian distribution $\mathcal{CN}\left(0,1/P_k\right)$. \fc{Each entry of the noise vector $\bm{w}(t)$ is an i.i.d. realization of $\mathcal{CN}(0,\sigma^2)$.} We denote the maximum delay and Doppler shift as $\tau_{\rm max}$ and $\nu_{\rm max}$. The bandwidth of the transmitted signal is $B$, assumed equal to the reciprocal of the symbol time. The symbol duration is given as $T_s=1/B$ while $T_f=n_pT_s$ is the time duration of $n_p$ symbols. In this work, we assume $\tau_{k,p} = i\, T_s $ for $i\in \{0,1, \dots, \tau_{\rm max}\}$ and $\nu_{k,p} = j\, \frac{1}{T_f}$ for $j\in \{-\nu_{\rm max}\!+\!1,\dots,\nu_{\rm max}\}$, i.e., that \ac{DD} shifts are integer multiples of $T_s$ and $1/T_f$. 
\fc{While fractional \ac{DD} shifts can be estimated through additional signal processing operations see e.g., ~\cite{2doffgrid} and reference therein, they can also be considered as multiple on-grid channel components at the price of reduced channel sparsity. The explicit incorporation of off-grid \ac{DD} shifts in our system model falls outside the scope of this paper and is left for future work.}
The decoder's task is to identify the unordered list $D(\bm{y})$ containing the $K_a$ messages transmitted by the active users. The \ac{PUPE} $P_e$ is defined as
\begin{equation}
P_e =  \frac{1}{K_a}\sum\limits_{k=1}^{Ka}\mathbb{P}[\bm{x}_k\notin D(\bm{y})]    \, .
\end{equation}

The system's performance is computed in terms of energy efficiency, and thus, the relevant metric is the minimum energy per bit $E_b/N_0 = n P / (\sigma^2 b)$ required to achieve a target error probability $P_e$.

\subsection{Transmitter}
\label{subsec:tx}
\begin{figure*}
    \centering
    \includegraphics[width=0.85\textwidth]{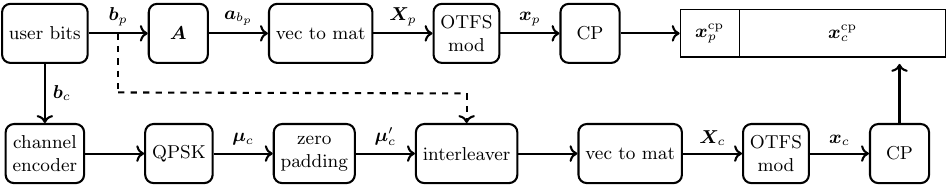}
    \caption{Transmission scheme}
    \label{fig:pipeline}
\end{figure*}

In this section, we describe the transmitter architecture with reference to Fig.~\ref{fig:pipeline}. 
The message to be transmitted is split in two parts of $b_p$ and $b_c$ bits each, so that $b=b_p+b_c$ and $b_p < b_c $. The vectors $\bm{b}_p\in \{0,1\}^{b_p}$ and $\bm{b}_c\in \{0,1\}^{b_c}$ refer to the two message parts chosen by a generic user. Following the strategies of~\cite{SIDMA} and~\cite{ODMA}, the two message parts are encoded with different procedures. The resulting coded symbols are then transmitted on successive orthogonal resources so that the two do not interfere with each other. 

The first $b_p$ bits are used to index a preamble of length $n_p$, whose time duration is ${T_f = n_p T_s}$. The mapping between bits and preambles is bijective. Each preamble symbol is i.i.d. and drawn from the distribution $\mathcal{CN}(0,1/n_p)$. The set of preambles constitutes the columns of the matrix $\bm{A}\in \mathbb{C}^{n_p \times 2^{b_p}}$, which is called \emph{sensing matrix}. The preamble set (hence the sensing matrix $\bm{A}$) is kept fixed once it has been designed, and is known to all users and the receiver. Once the transmitting user has chosen the preamble, such preamble is used to fill column-wise the matrix $\bm{X}_p \in \mathbb{C}^{M_p\times N_p}$ where $M_p N_p = n_p$. As the symbols in $\bm{X}_p$ are in the \ac{DD} domain, the \ac{OTFS} modulator block converts them to the time domain. The modulator is implemented using the \ac{IDZT} as
\begin{equation}
\label{eq:IDZT_mat}
        \bm{x}_p =\mathrm{vec}(\bm{X}_p\bm{F}_{N_p}^H) \, ,
\end{equation}
where the $\mathrm{vec}(\cdot{})$ operator vectorizes the argument matrix along its columns, $\bm{F}_N$ is the \ac{DFT} matrix of size $N$, and $^H$ is the Hermitian operator. The last operation concerning the preamble is the addition of a \ac{CP} of length $\tau_{\rm max}$ in front of it. The result of the aforementioned processing steps is the vector $\bm{x}_p^{cp}$ of length $n_p + \tau_{\rm max}$ symbols. 

The remaining $b_c$ bits are processed by the blocks at the bottom of Fig.~\ref{fig:pipeline}. First, the bit sequence $\bm{b_c}$ is encoded with the \ac{CRC}-aided polar encoder according to the \ac{5G-NR} standard. 
The encoded bits are modulated using \ac{QPSK}, and the resulting vector $\bm{\mu}_c$ is padded with zeros to obtain $\bm{\mu}_c'$ of length $n_c$. At this point, an interleaver $\bm{\Pi}_i$ performs a pseudo-random permutation of $\bm{\mu}_c'$, which is uniquely determined by the seed~$\bm{b}_p$. By doing so, the preamble is used as side information for the data encoder and will be exploited at the receiver to reverse the interleaving operation. The rationale of this operation is to minimize the \ac{MUI} given that the interleaving patterns are independently chosen by each user. The interleaved vector $\bm{\Pi}_{\bm{b}_{p}}(\bm{\mu}_c')$ is then reshaped to fill column-wise the \ac{DD} matrix $\bm{X}_c \in \mathbb{C}^{M_c\times N_c}$ with $M_c N_c = n_c$. The OTFS modulator converts $\bm{X}_c$ into its discrete time representation  $\bm{x}_c=\mathrm{vec}(\bm{X}_c\bm{F}_{N_c}^H)$, and finally, a \ac{CP} of length $\tau_{\rm max}$ is prepended to $\bm{x}_c$ to obtain $\bm{x}_c^{cp}$. 

The baseband representation of the overall transmitted signal by one of the active users is given by $\bm{x}=[\bm{x}_p^{cp},\bm{x}_c^{cp}]\in \mathbb{C}^n$. The total number of complex degrees of freedom of $\bm{x}$ is given by $n = n_b + n_c + 2\tau_{\rm max}$.

\subsection{Receiver}
The $K_a$ active users concurrently transmit over the same resources. Nevertheless, as shown in Eq.~\eqref{eq:channel}, the transmitted signals face different channels and will arrive at the receiver with different \ac{DD} offsets, multiplicity, and complex amplitude gains. To retrieve the transmitted messages, the receiver needs to detect the number of active users and estimate their respective channels so that the received signals can be equalized and decoded. The preamble and data bits are estimated sequentially using different techniques. The continuous-time signal at the output of the channel is converted to baseband and sampled to obtain $\bm{y}$ as in Eq.~\eqref{eq:channel}. The receiver extracts the vectors $\bm{y}_p^{cp}$ and $\bm{y}_c^{cp}$ from $\bm{y}$ by taking the first $n_p+\tau_{\rm max}$ and the subsequent $n_c+\tau_{\rm max}$ complex symbols.

\subsubsection{\Ac{CS} preamble decoder}
\label{subsec:phase1}
From the signal $\bm{y}_p^{cp}$, we remove the \ac{CP} to obtain $\bm{y}_p$. The preamble decoder has the objective to estimate the active preamble set, denoted as $\mathcal{L}_p$, together with the channel parameters of all active users. This task can be seen through the lens of \ac{CS}: the receiver must understand which columns of the sensing matrix are active by observing a noisy version of their superposition. In the case of \ac{GMAC} or static fading, the sensing matrix observed by the receiver is exactly $\bm{A}$. When we consider a doubly-selective channel, instead, each preamble undergoes a channel with different delays and Doppler shifts. Their effect can be directly included in $\bm{A}$ by expanding the columns and including all possible shifted versions of each preamble.
In the expanded sensing matrix, denoted as $\bm{A}_{\rm exp}$ and of dimension $n_p\times \left( 2^{b_p}\,(\tau_{\rm max}+1)\,2\nu_{\rm max}\right)$, each preamble will be found multiple times, once for each \ac{DD} combination. The column of $\bm{A}_{\rm exp}$ related to the $i$-th preamble and a $(\tau,\nu)$ \ac{DD} shift is 
\begin{equation}
    \label{eq:sigma}
    \bm{a}_{i}^{\tau,\nu} = (\textbf{F}_{N_p} \otimes \textbf{I}_{M_p})(\mathrm{e}^{-j\frac{2\pi}{N_pM_p}\nu\tau}\bm{\Delta}^{\nu}\mypi^{\tau})(\textbf{F}_{N_p}^H \otimes \textbf{I}_{M_p})\bm{a}_i \, ,
\end{equation}
where $\textbf{I}_{N}$ is the $N \times N$ identity matrix, $\otimes$ is the Kronecker product, ${\boldsymbol{\Delta}^{k_i} = \diag{}[\ee{}^{\frac{j2\pi}{N_pM_p}\nu\cdot(0)}, ... \,, \ee{}^{\frac{j2\pi}{N_pM_p}\nu\cdot(N_pM_p-1)} ]}$ and $\mypi$ is the cyclic shift matrix of size $N_pM_p \times N_pM_p$
\begin{equation}
       \mypi = \begin{bmatrix}
    0 & \cdots & 0 & 1\\
    1 & \cdots & 0 & 0\\
    \vdots & \ddots & \vdots & \vdots \\
    0 & \cdots & 1 & 0
\end{bmatrix} \, .
\end{equation}
The vector $\bm{a}_{i}^{\tau,d}$ is the vectorized \ac{DD} signal seen at the receiver after the transmission of the preamble $\bm{a}_{i}$ over a single tap channel with delay $\tau$, Doppler $\nu$ and unitary gain $h=1$. 

Many algorithms are available in the literature to deal with the considered noisy \ac{CS} decoding problem. Here we resort to \ac{AMP}~\cite{AMP}. 
The input of the \ac{CS} decoder is the measurement vector $\bm{y}_p^{dd} = (\textbf{F}_{N_p} \otimes \textbf{I}_{M_p})\bm{y_p}$, which represents the vectorized \ac{DD} preamble signal observed by the receiver. Considering only the preamble part, the input-output relation in Eq.~\eqref{eq:channel} can be rewritten as
\begin{equation}
    \bm{y}_p^{dd}=\bm{A}_{\rm exp}\,\bm{\varphi} + \bm{w}^{dd} \, ,
\end{equation}
where $\bm{\varphi} \in \mathbb{C}^{( 2^{b_p}\,(\tau_{\rm max}+1)\,2\nu_{\rm max}) \times1}$ represents the sparse vector that the receiver aims to estimate and $\bm{w}^{dd}$ is the noise term in \ac{DD} domain. In particular, the vector $\bm{\varphi}$ has $\sum_{k=1}^{K_a}P_k$ non-zero entries whose values are given by the $h_{k,p}$ channel coefficients. By using the decoding algorithm \ac{AMP}, we obtain the estimate $\hat{\bm{\varphi}}$ of $\bm{\varphi}$ starting from $\bm{y}_p^{dd}$ and $\bm{A}_{\rm exp}$. Inspecting $\hat{\bm{\varphi}}$, we retrieve the estimated set of active preambles $\hat{\mathcal{L}}_p$ with their respective channel parameters composed of the number of propagation paths $\hat{P}_k$ and the complex attenuation, delay, and Doppler shift $(\hat{h}, \hat{\tau}, \hat{\nu})$ of each path. We call $\mathcal{\hat{H}}$ the set of estimated parameters describing the channels of all users, i.e., the $\hat{P}_k$ triplets of $(\hat{h}, \hat{\tau}, \hat{\nu})$. Depending on the prior statistical knowledge of $\bm{\varphi}$ available at the receiver, different denoisers can be adopted to enhance the performance of the \ac{AMP} decoder~\cite{liu18}. Note that for small $b_p$ and large active user population size~$K_a$, the probability that two users select the same preamble is non-negligible. We discuss this issue in Sec.~\ref{sec:numres}.

\begin{figure}
    \centering
    \includegraphics[width=1\linewidth]{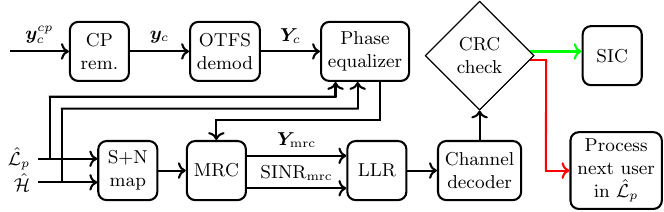}
    \caption{Receiver scheme for data decoding.}
    \label{fig:phase2RX}
    \vspace{-0.2cm}
\end{figure}

\subsubsection{Data decoder}
The operations performed to estimate the vector $\bm{b}_c$ of each active user are shown in Fig.~\ref{fig:phase2RX}. After the \ac{CP} removal, the \ac{DD} representation of the signal $\bm{y}_c$ is computed as
\begin{equation}
\label{eq:DZT_mat}
        \bm{Y}_c =\textbf{F}_{N_c} \mathrm{vec}^{-1}_{M_cN_c}(\bm{y}_c) \,,
\end{equation}
where $\mathrm{vec}^{-1}_{ij}(\cdot{})$ reshapes the argument vector of length $(i\cdot j)$ into a matrix of size $i \times j$, filling the matrix column-wise. The operation in Eq.~\refeq{eq:DZT_mat} is known as \ac{DZT} and is performed by the \ac{OTFS} demodulator block. The matrix $\bm{Y}_c$ contains multiple \ac{DD} repetitions of the interleaved source signal transmitted by each of the $K_a$ users, where each repetition is shifted and scaled according to the channel parameters of the considered user. 

At this point, the receiver employs a single-user decoding strategy, treating the \ac{MUI} and \ac{ISI} as Gaussian noise and triggering the \ac{SIC} procedure whenever decoding succeeds. Exploiting the knowledge gained from the \ac{CS} decoding phase (see Sec.~\ref{subsec:phase1}), related to the active interleavers and their relative channel parameters, 
the receiver can estimate the aggregate signal strength on each \ac{DD} symbol in the \ac{OTFS} matrix $\bm{Y}_c$. By including also the noise power a signal plus noise map is constructed. The receiver's strategy is to use \ac{MRC}~\cite{Brennan59} at the symbol level among the channel echoes before attempting the decoding on the codeword. In particular, it extracts the selected interleaver from $\hat{\mathcal{L}}_p$ and the the corresponding channel parameters from $\hat{\mathcal{H}}$. It then proceeds by recovering from $\bm{Y}_c$ the symbols corresponding to each multipath echo with the aim of coherently combining them to get an enhanced \ac{SINR}. 
%
%
Note that, before combining, the symbols corresponding to each echo in $\bm{Y}_c$ have to be shifted back, de-interleaved, and equalized since different \ac{DD} shifts introduce systematic phase rotations. These phase rotations can be obtained by means of the \ac{DZT}\cite{OTFSbook}, by computing the discrete input-output relation in \ac{DD} domain. This can be done by considering a simple single user channel with a unique propagation path with \ac{DD} coordinates $(\tau,\nu)$ and unitary gain:
\begin{equation}
\bm{Y}[m,n]  =  \ee{}^{\frac{j 2\pi }{MN}\nu (m - \tau)} \bm{X}[ (m - \tau)_{M},(n-\nu)_{N}] \ee{}^{\frac{j 2\pi }{N}(n - \nu) \left\lfloor \frac{m - \tau}{M} \right\rfloor}
\label{eq:phase}
\end{equation}
where $\bm{X}\in\mathbb{C}^{M\times N}$ denotes the \ac{DD} representation of the transmitted signal. With the help of the signal plus noise map and the estimated channel coefficients, the receiver can create the combined observation of the codeword $\bm{Y}_{\rm mrc}$. This is obtained as the symbol-level weighted sum of the phase-corrected echoes according to their respective \ac{SINR}s. Finally, the combined observation $\bm{Y}_{\rm mrc}$ and the corresponding \ac{SINR} values are used to produce the codeword bits \acp{LLR} that are given in input to the \ac{SCL} Polar decoder~\cite{SCL}.
The output of the channel decoder is considered correct whenever the \ac{CRC} check is successful.
In this case, the receiver triggers the \ac{SIC} procedure and cancels all the detected echoes of the decoded signal from $\bm{Y}_c$. %
The iterative process continues until either all users in $\hat{\mathcal{L}}_p$ are successfully decoded, or until the decoder is no longer able to decode any remaining message. 


\section{Numerical results}
\label{sec:numres}
\begin{figure}
    \centering
    \includegraphics[width=0.87\linewidth]{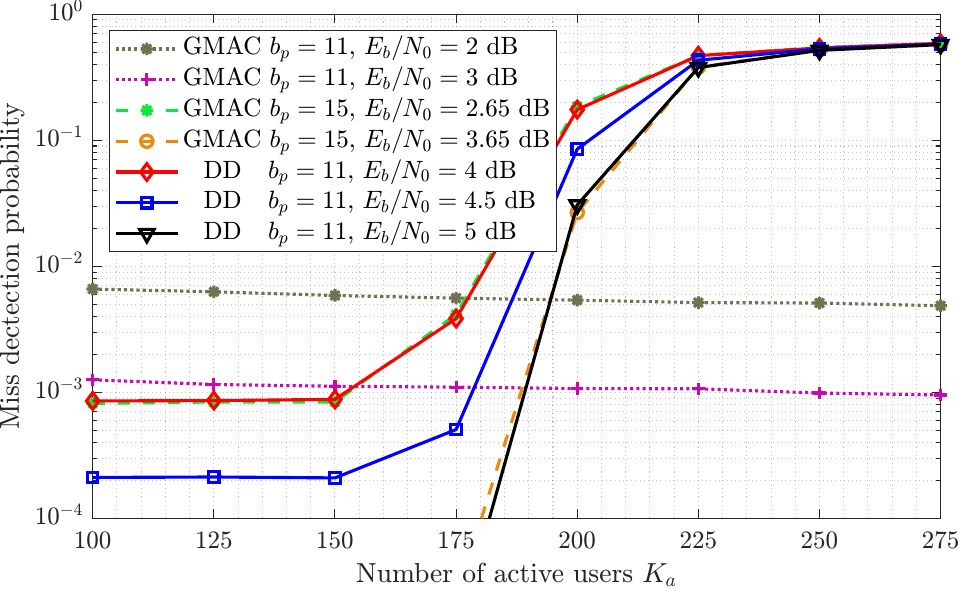}
    \caption{Preamble miss detection probability versus the number of active users for different system and channel parameters.}
    \label{fig:phase1}
    \vspace{-0.2cm}
\end{figure}
In Fig.~\ref{fig:phase1}, we show the performance of the \ac{AMP} decoder in terms of miss detection probability, namely the ratio of undetected preambles over the transmitted ones, for a varying number of active users $K_a$, assuming a single-path channel with unitary channel coefficient. We dedicate to the preamble transmission $n_p=640$ complex channel uses, organized in an \ac{OTFS} matrix with $M_p = 40$ and $N_p = 16$. Other system parameters are $b_p = 11$, $\tau_{\rm max}=3$ and $\nu_{\rm max} = 2$, hence, the expanded sensing matrix at the receiver  $\bm{A}_{\rm exp}$ has dimensions $640 \times 2^{15}$. A preamble is considered undetected also if the receiver does not exactly recover its \ac{DD} shift. 
A high recovery probability in the first phase is key for the overall system to work successfully as it provides key information for the data decoding phase. 
With the selected parameters, represented with the solid curves in Fig.~\ref{fig:phase1}, the recovery probability does not limit the overall system performance up to $K_a = 175$, having in mind a target $P_e=0.05$. Other results are shown in the same figure for the case where $\tau_{\rm max}=0$ and $\nu_{\rm max} = 0$, thus effectively resulting in the \ac{GMAC} scenario. In this latter setting, it is possible to observe a smaller required $E_b/N_0$ to achieve similar error probabilities thanks to the less severe undersampling ratio (dotted curves). On the other hand, maintaining the same \ac{SNR} and target miss detection probability, a greater number of bits could be embedded in the preamble signal (dashed curves). This phenomenon also demonstrates 
that $\bm{A}_{\rm exp}$ behaves as a true random i.i.d. matrix with Gaussian entries~\cite{AMP}. 

\begin{figure}
    \centering
    \includegraphics[width=0.87\linewidth]{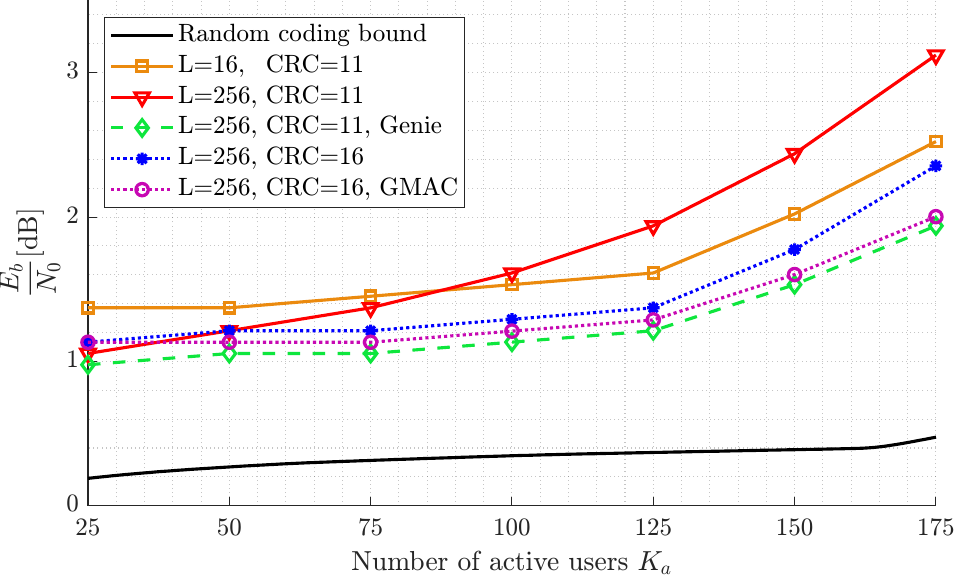}
    \caption{Overall $E_b/N_0$ required to achieve $P_e = 0.05$ as a function of the number of active users assuming a single-path channel with unitary gain.}
    \label{fig:phase2}
    \vspace{-0.2cm}
\end{figure}
A first result related to the second protocol phase is shown in Fig.~\ref{fig:phase2}, where the required $E_b/N_0$ needed to achieve a target $P_e=0.05$ is shown. 
Ideal decoding of the first phase is assumed, so that $\hat{\mathcal{L}}_p=\mathcal{L}_p$ and $\hat{\mathcal{H}}=\mathcal{H}$. \fc{The energy overhead due to the first protocol phase is already included in Fig.~\ref{fig:phase2}, and it is equal to an $E_{b}/N_{0}$ of 3 dB for the \ac{GMAC} case and of 4 dB for all others. The overall $E_{b}/N_{0}$ is computed as the average, weighted according to number of information bits contained in each phase, between the preamble $E_{b}/N_{0}$ and the one required by the second protocol phase to reach the target $P_e$. The \acp{CP} are not counted into the energy budget.}
The additional simulation parameters are $M_c = 115$, $N_c = 128$, $b_c = 89$ and the considered channel is the same as the one with \ac{DD} shifts in the previous experiment. The employed polar code has block-length equal to 512 and it takes as input $\bm{b}_c$ together with additional \ac{CRC} bits. Here we describe the tradeoff between list size and \ac{CRC} length. Large values of the list size enhance the error correction performance of the code. However, larger lists also lead to more undetected errors that can only be compensated with stronger \acp{CRC}. In Fig.~\ref{fig:phase2}, we report the performance of the scheme when considering list sizes $16$, $256$, and $256$ with ideal error detection (denoted with the \textit{Genie} label). We see that, for the case $L=256$, better performance is achieved when the number of users is low, while the $L=16$ case shows better performance when the number of active users increases, i.e., for $K_a\geq125$. This phenomenon can be explained by the fact that undetected errors are particularly detrimental when \ac{SIC} is employed. In fact, if a codeword is erroneously marked as correctly decoded, \ac{SIC} is triggered, causing the introduction of additional interference that the receiver will not be able to remove later on. 
To enable the use of bigger lists, it is necessary to increase the error detection capability of the overall code by increasing the number of CRC bits. The performance of the scheme with $L=256$ and a longer 16-bit \ac{CRC} are also shown in Fig.~\ref{fig:phase2}. \fc{Thanks to the longer \ac{CRC}, the undetected errors are strongly reduced. Nevertheless, it is worth noting that even this minor increase in code rate 
can lead to reduced error protection, which is the limiting factor for small $K_a$. Both dotted curves in Fig.~\ref{fig:phase2} adopt the same polar encoder. Still, the \ac{GMAC} performance are better due to the reduced preamble energy penalty and a slightly lower code rate due to $b_c=85$.}
\begin{figure}
    \centering
    \includegraphics[width=0.87\linewidth]{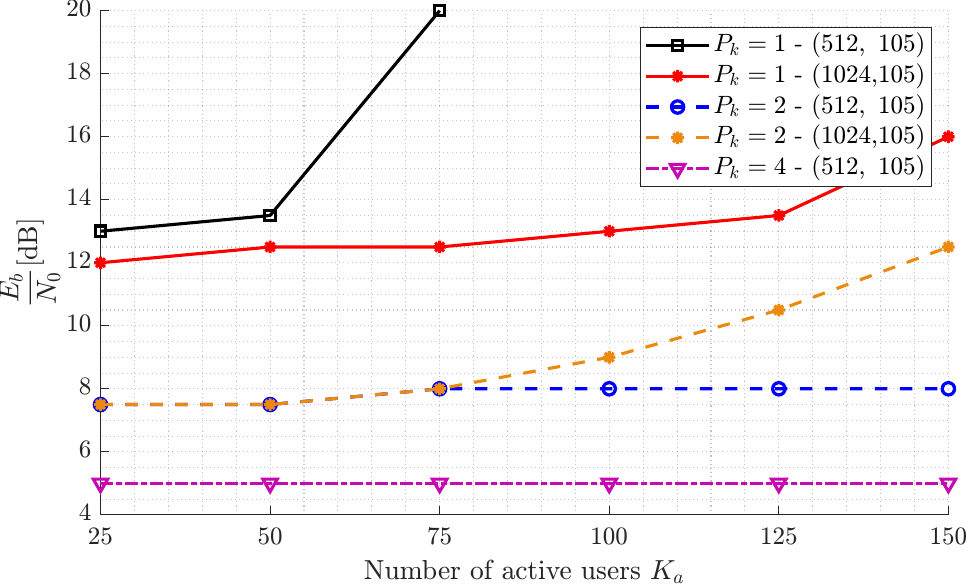}
    \caption{$E_b/N_0$ needed in the the second protocol phase to achieve $P_e = 0.05$ in the presence of fading as a function of the number of active users for a varying number of multipath components $P_k$.}
    \label{fig:phase2Fading}
    \vspace{-0.2cm}
\end{figure}

In Fig.~\ref{fig:phase2Fading}, the protocol performance is shown in the presence of the fading model described in Sec.~\ref{sec:sys_mod}. We consider genie-aided error detection, meaning that no undetected errors occur. Assuming $P_k=1$ for all active users, the scheme with the (512,105) code requires large $E_b/N_0$ values due to the too high outage probability, meaning that when the magnitude of the channel gain is too small, the decoder is not able to retrieve the transmitted codeword even in presence of little \ac{MUI}. A reduction in code rate enables a much better scaling of the scheme for a growing number of active users. When instead considering, $P_k\ge2$ for each user, the outage probability becomes much smaller, and the \ac{MRC} enables the exploitation of \emph{multi-path diversity} thus rendering feasible the application of the higher rate polar code. This effect demonstrates the potential gain related to the capability of resolving and recombining the multipath echoes. \fc{Overall, while a growing number of multipath components renders the preamble estimation more difficult, it could result in better achievable performance of the scheme over fading channels.}

In each of the protocol phases, we generate a signal in \ac{DD} and then we convert it to the time domain by means of the \ac{OTFS} modulator. Processing the signals in the \ac{DD} domain allows us to cope with the double selectivity of the considered channel. When the channel \ac{DD} spread is well confined inside an \ac{OTFS} matrix, satisfying the crystallization condition~\cite{OTFSbits1},
%
we have that all symbols within the \ac{OTFS} matrix will experience the same channel up to some phase rotations that can be easily accounted for, see Eq.~\eqref{eq:phase}. 
This stationary behavior is in stark contrast with other multi-carrier schemes like \ac{OFDM}, where each subcarrier experiences different 
gains and \acp{SNR}. 

For the two-step approach, some discussion related to the use of the \ac{DD} channel estimate among the different protocol phases should be made. In the proposed scheme, we exploit the preamble sequences, transmitted for each user within $\bm{X}_p$, to estimate the channels. The obtained channel estimate is then exploited to equalize the signal $\bm{Y}_c$ related to the second protocol phase. 
\fc{In the \ac{OTFS} literature, channel estimation is typically done on a per-frame basis. Nevertheless, it is important to note that when two \ac{DD} signals are transmitted sequentially, the channel variations acting on the two are not arbitrary. In particular, the \ac{DD} channel representation enables a geometrical interpretation of the scene, where the \ac{DD} coordinates of each path are related to the position and speed of a specific scatterer present in the scene. 
These attributes are, in general, slowly changing and could be considered approximately constant for a long time interval~\cite{OTFS_original,OTFSbook,2011matz_book}. 
In this study, we consider a constant \ac{DD} channel representation that spans both phases. Nevertheless, it is worth noting that, due to the presence of Doppler shifts, the considered input-output relation in Eq.~\eqref{eq:channel} has the effect of a fast-varying frequency-selective channel, and is thus more general than the commonly adopted quasi-static fading model. }

To avoid interference between $\bm{x}_p$ and $\bm{x}_c$, a \ac{CP} is used in front of the latter signal. This choice, together with the fact that the two signals have different lengths and thus different sizes of the \ac{DD} matrices, requires to apply some minor adjustments of the channel estimates derived in the first phase when exploited in the second one. While the delay resolution is proportional to the bandwidth $B$, which remains unchanged between the two phases, the Doppler resolution is proportional to the signal duration of each phase, and this can be easily accounted for. 
The time duration of $\bm{x}_c$ equals $\alpha T_f$ with $\alpha \in \mathbb{R}$, and thus when the receiver estimates a Doppler shift of $\nu_p$ bins within the \ac{DD} preamble matrix, that same shift will be of $\nu_c=\alpha \nu_p$ in the following \ac{OTFS} matrix. In general we have that $\frac{\nu_p}{T_f}=\frac{\nu_c}{\alpha T_f}$ and the system can be designed so that $\alpha \in \mathbb{N}$. An additional adjustment has to be applied to each complex gain $h_{k,p}$ to account for the total phase rotation that accumulates during the \ac{CP} transmission due to the potential presence of a Doppler contribution. 

In~\cite{mirri2024amp} an ad-hoc \ac{AMP} denoiser for fractional \ac{DD} channel has been recently developed, showing promising performance even in this more challenging setting. Note that fractional component can be seen as multiple integer ones, and thus the \ac{MRC} principle can still be effectively used. Therefore, with some minor adjustment, the proposed scheme could also be used in the non-idealized scenario of fractional \ac{DD} shifts.

Depending on the number of bits dedicated to the preamble and the number of active users $K_a$, the probability of preamble collisions may become non-negligible. \fc{Differently than in the \ac{GMAC} setting, the user-specific channel effect, such as \ac{DD} shifts and related power profile, could be exploited to separate the contributions of various users. In general, relying heavily on such an approach might not always be effective since colliding users might have with too similar or indistinguishable channels.}  
In general, it would be better to avoid preamble collision in the first place by increasing $b_p$. The limit of this latter solution is the convergence of the \ac{CS} decoder, which, in the case of \ac{AMP}, can be predicted by state evolution~\cite{AMP}. There is thus a tension between the number of preambles, driven by $b_p$, and the maximum channel \ac{DD} spread that we are able to support. 
To mitigate this phenomenon one could increase the preamble length at the cost of reduced number of channel uses for the second protocol phase. Therefore, this latter choice trades a growing preamble set with for higher \ac{MUI} in the second protocol phase. 


\section{Conclusions}
\label{sec:conc}
\fc{In this paper, we consider the problem of \ac{UMAC} in the presence of doubly-dispersive channels. We propose and discuss the use of \ac{DD} signal transmission based on multi-frame \ac{OTFS} modulation as a key component for the construction of a grant-free multi-user scheme. Several design aspects are discussed and numerical results demonstrates that the proposed scheme shows competitive performance even in the presence of 
\ac{DD} shifts. Furthermore, in the presence of multiple propagation paths, multipath diversity can be effectively exploited to combat the impact of fading. Further studies are needed to fully assess the scheme capabilities by considering the impact of imperfect channel estimation, fractional shifts and preamble collisions.}

\bibliographystyle{IEEEtran}
\bibliography{bibtexs/IEEEabrv,bibtexs/OTFS}

\begin{acronym}
        \acro{2SRA}{two-step random access}
        \acro{4-QAM}{4-quadrature amplitude modulation} 
        \acro{16-QAM}{16-quadrature amplitude modulation}

        \acro{6G}{sixth generation}
        \acro{4SRA}{four-step random access}
        \acro{5G-NR}{5G new radio}
        \acro{ADC}{analog-to-digital converter}
        \acro{AMP}{approximate message passing}
        \acro{AWGN}{additive white Gaussian noise}
        \acro{BER}{bit error rate}
        \acro{BLER}{block error rate}
        \acro{BP}{belief propagation}
        \acro{BS}{base station}
        \acro{CCS}{coded compressed sensing}
        \acro{CAZAC}{constant amplitude zero autocorrelation}
        \acro{c.c.u.}{complex channel use}
        \acro{CP}{cyclic prefix}
        \acro{CRC}{cyclic redundancy check}
        \acro{CRDSA}{contention resolution diversity slotted Aloha}
        \acro{CS}{compressed sensing}
        \acro{CSA}{coded slotted Aloha}
        \acro{c.u.}{channel use}
        \acro{DAC}{digital-to-analog converter}
        \acro{DD}{delay-Doppler}
        \acro{DFT}{discrete Fourier transform}
        \acro{DZT}{discrete Zak transform}
        \acro{EP}{embedded pilot}
        \acro{GMAC}{gaussian multiple access channel}
        \acro{ICI}{inter-carrier interference}
        \acro{ICSI}{ideal channel state information}
        \acro{IDMA}{interleaver division multiple access}
        \acro{IDFT}{inverse discrete Fourier transform}
        \acro{IDZT}{inverse discrete Zak transform}
        \acro{IFFT}{inverse fast Fourier transform}
        \acro{IoT}{Internet of Things}
        \acro{IRSA}{irregular repetition slotted Aloha}
        \acro{ISI}{inter-symbol interference}
	    \acro{LDPC}{low-density parity-check}
        \acro{LEO}{low-Earth orbit}
        \acro{LLR}{log-likelihood ratio}
        \acro{LMMSE}{linear minimum mean squared error}
        \acro{LTE}{Long Term Evolution}
        \acro{MAC}{multiple access}
        \acro{MIMO}{multiple-input multiple-output}
        \acro{MMSE}{minimum mean square error}
        \acro{MPR}{multi-packet reception}
        \acro{MRC}{maximal-ratio combining}
        \acro{MTC}{machine-type communication}
        \acro{MMTC}{massive machine-type communication}
        \acro{MRA}{massive random access}
        \acro{MTO}{many-to-one}
        \acro{MUI}{multi-user interference}
        \acro{NB-IoT}{Narrowband IoT}
        \acro{NMSE}{normalized mean squared error}
        \acro{NOMA}{non-orthogonal multiple access}
        \acro{NR}{new radio}
        \acro{ODMA}{on-off division multiple access}
        \acro{OFDM}{orthogonal frequency-division multiplexing}
        \acro{OMP}{orthogonal matching pursuit}
        \acro{OTFS}{orthogonal time frequency space}
        \acro{OTO}{one-to-one}
        \acro{PAM}{pulse-amplitude modulation}
        \acro{PAPR}{peak-to-average power ratio}
        \acro{PBCH}{physical broadcast channel}
        \acro{PDSCH}{physical downlink shared channel}
        \acro{PO}{PUSCH occasion}
        \acro{PRACH}{physical random access channel}
        \acro{PRB}{physical resource block}
        \acro{PSS}{primary synchronization signal}
        \acro{PUPE}{per-user probability of error}
        \acro{PUSCH}{physical uplink shared channel}
        \acro{QPSK}{quadrature phase shift keying}
        \acro{QAM}{quadrature amplitude modulation}

        \acro{RA}{random access}
        \acro{RCP}{reduced cyclic prefix}
        \acro{RCU}{random coding union}
        \acro{S-IDMA}{sparse-interleaver-division multiple access}
        \acro{SB-IDMA}{sparse block interleaver division multiple access}
        \acro{SCL}{successive cancellation list}
        \acro{SCS}{sub-carrier spacing}
        \acro{SE}{spectral efficiency}
        \acro{SIC}{successive interference cancellation}
        \acro{SINR}{signal-to-interference-plus-noise ratio}
        \acro{SNR}{signal-to-noise ratio}
        \acro{SPARC}{sparse regression code}
        \acro{S1D}{superimposed pilot}
        \acro{SSS}{secondary synchronization signal}
        \acro{TA}{time advance}
        \acro{TF}{time-frequency}
        \acro{TBS}{transport block size}
        \acro{TIN}{treat-interference-as-noise}
        \acro{UE}{undetected error}
        \acro{UT}{user terminal}
        \acro{UMAC}{unsourced multiple access}
        \acro{ZC}{Zadoff-Chu}
        \acro{ZF}{zero forcing}
        \acro{ZP}{zero padding}
\end{acronym}

\end{document}